 \documentclass [12pt,a4paper]{article}
\usepackage{times}

\DeclareFontFamily{OT1}{times}{}
\DeclareFontShape {OT1}{times}{m }{n }{ <-> ptmr }{}
\DeclareFontShape {OT1}{times}{bx}{n }{ <-> ptmb }{}
\DeclareFontShape {OT1}{times}{m }{it}{ <-> ptmri}{}
\DeclareFontShape {OT1}{times}{bx}{it}{ <-> ptmbi}{}
\usepackage{amsmath}
\usepackage{amsfonts}
\usepackage{amssymb}
\usepackage{latexsym}
\newcommand{\bpi}{\boldsymbol{\pi}}
\newcommand{\bphi}{\boldsymbol{\phi}}
\newcommand{\bfeta}{\boldsymbol{\eta}} 

\newcommand{\BRA}{\langle\kern -.2em\langle} 
\newcommand{\KET}{\rangle\kern -.2em\rangle} 
\newcommand{\FOU}{\widehat}           
\newcommand{\HUS}{\text{H}} 

\numberwithin{equation}{section}               

\begin{document}

\title{\bf\vspace{-3.5cm} Nonlinear generalized functions and
                          the Heisenberg-Pauli foundations of\\
                          Quantum Field Theory}

\author{ {\bf Jean-Fran\c{c}ois Colombeau}\\
             (jf.colombeau@wanadoo.fr)\\
         {\bf Andre Gsponer}\\
         {\it Independent Scientific Research Institute, Oxford OX4 4YS, UK}\\
         {\bf Bernard Perrot}\\
         {\it ENSEIRB, Universit\'e de Bordeaux 1, 33405 Talence, France}\\
       }


\maketitle

\begin{abstract}
In 1929 Heisenberg and Pauli laid the foundations of QFT by quantizing the fields (method of canonical quantization). This general theory of quantized fields has remained undisputed up to now. We show how the unmodified Heisenberg-Pauli calculations make sense mathematically by using a theory of generalized functions adapted to nonlinear operations. By providing an appropriate mathematical setting, nonlinear generalized functions open doors for their understanding but there remains  presumably very hard technical problems. (i) Domains of the interacting field operators: a priori the H-P calculations give time dependent dense domains, what is not very convenient; (ii) Calculations of the resulting matrix elements of the $\mathbf{S}$ operator: from the unitarity of the $\mathbf{S}$ operator as a whole there are no longer ``infinities,'' but a priori there is no other hope than heavy computer calculations; (iii) Connection with renormalization theory: it should provide an approximation when the coupling constant is small. The aim of this paper is to present, on the grounds of a standard mathematical model of QFT (a self interacting scalar boson field), a basis for improvement without significant prerequisites in mathematics and physics. It is an attempt to use nonlinear generalized functions in QFT, starting directly from the calculations made by physicists, in the same way as they have already been used in classical mechanics and general relativity.  

\end{abstract}

\noindent \emph{Category}: MP, mathematical physics.
\noindent \emph{Comments}: 20 pages, research-expository paper.

\section{Prerequisites on generalized functions}
\label{gen:0}

As is well known the Heisenberg-Pauli calculations (or ``\emph{canonical formalism},'' see \cite[p.\,292]{W}), require a concept of generalized functions adapted to nonlinear calculations (and whose values are unbounded operators on a Hilbert space). These generalized functions were introduced 25 years ago, thanks to support of L.~Nachbin \cite{C5,C6} and L.~Schwartz \cite{C7}, but they suffered from the widespread belief that ``there will never exist a theory of generalized functions adapted to nonlinear calculations.'' 

For the theory and applications of scalar valued nonlinear generalized functions, see the recent introductions in \cite{C1, C2, G}, the survey \cite{C3} and the book \cite{C8} on their use in \emph{continuum mechanics}, and the survey \cite{S-V} and the book \cite{G-K-O-S} on their use in \emph{general relativity}.  The simplest definition of nonlinear generalized functions can be found in \cite{S-V} and references therein.  The best and most recent textbook is \cite{G-K-O-S}.  

One calculates with generalized functions as with $\mathcal{C}^\infty$ functions (although they seem basically very different, in particular when one attempts a physical interpretation): if $\Omega$ is an open set in $\mathbb{R}^N$ \emph{the generalized functions on $\Omega$ are treated as if they were familiar $\mathcal{C}^\infty$ functions that can be differentiated and multiplied freely, but there is a novelty.} Classically if $\HUS$ denotes the Heaviside function it is natural to state $\HUS^2=\HUS$. In the present context this is false: one has $\HUS^2 \neq \HUS$. 

Let us explain this basic point in detail. In physics the Heaviside function $\HUS(x)$ represents a function whose values jump from 0 to 1 in a tiny interval of length $\epsilon$ around $x=0$. Thus it is obvious that $\int (\HUS^2(x) - \HUS(x) ). \varphi(x)~dx$ tends to 0 when $\epsilon \rightarrow 0^+$ if $\varphi$ is a bounded function. But since $\HUS'$ is unbounded one has $\int (\HUS^2(x) - \HUS(x) ).  \HUS'(x)~dx = 1/3 - 1/2 = -1/6$, as obvious from elementary calculations. This shows that one is not allowed to state $\HUS^2=\HUS$ in a context where the function $\HUS^2-\HUS$ could be multiplied by a function taking infinite values such as the Dirac function $\delta=\HUS'$.  Therefore, in the $\mathcal{G}$-context,\footnote{For brevity, we write ``$\mathcal{G}$-context'' to refer to the nonlinear generalized functions context or setting.} one has to distinguish between ``infinitesimally nonzero'' functions such as $\HUS^2-\HUS$ and the genuine zero function, because ``infinitesimally nonzero quantities'' multiplied by ``infinitely large quantities'' can give significant nonzero results.  For this reason we introduce a new notation: for two generalized functions $G_1$ and $G_2$ on an open set $\Omega$ we write $G_1 \approx G_2$ (``$G_1$ \emph{associated} to $G_2$'') iff $\forall \varphi \in \mathcal{C}_0^\infty(\Omega)$ (i.e., infinitely differentiable with compact support in $\Omega$) the number $\int_\Omega  (G_1-G_2)(x)\,\varphi(x)~dx$ is ``infinitesimal'' (i.e., depends on a small parameter $\epsilon>0$ and tends to 0 as $\epsilon \rightarrow 0^+$). 

The above was at the origin of widespread claims from mathematicians that  ``multiplication of generalized functions is impossible.''  Not only is it not useful and obviously impossible to keep classical equalities such as $\HUS^2=\HUS$, but also physics displays instances in which several different Heaviside functions (associated to each other) are requested to model different physical variables: see the example of elastoplastic shock waves \cite[fig.~6]{C3}, \cite[p.\,106]{C8}, \cite[p.\,185--189]{Bia}, \cite{C9,LR}. From ``concrete'' examples in continuum mechanics \cite{C1, C2, C3, C4, C8, Bia, LR} the basic mathematical idea became obvious: classical analysis was built at an epoch in which physics did not show ``ambiguous products of distributions.'' \emph{The ``ambiguities in products of distributions'' that were found more recently in physics} (in classical continuum mechanics \cite[and references therein]{C1,C2,C3,C8,LR}, in general relativity \cite[and references therein]{G-K-O-S, S-V}) \emph{appear to be due to an excessively strong mathematical idealization of physical objects which destroys (or at least does not contain) the information needed to solve these problems.} It is shown on physical examples in \cite{C1, C2, C3, C8} how nonlinear generalized functions contain more information. Indeed it was found that they were mathematically compelled to contain more information in order to solve the problem of multiplication of distributions \cite{C5, C6, C7}, even before their use in physics began. 

To summarize one calculates with nonlinear generalized functions as in classical calculations with infinitely differentiable functions (using =), but a basic novelty lies in that \emph{the classical equality of functions splits into $=$ (the ``true'' or ``strong'' or ``algebraic'' equality) and the association $\approx$ (also called ``weak'' equality)}. Concerning association the basic point is that $G_1 \approx G_2$ \emph{does not imply automatically} $G.G_1 \approx G.G_2$ if $G$ is another generalized function. But $G_1 \approx G_2$ implies $\partial/\partial_{x}G_1 \approx \partial/\partial_{x}G_2$.  For example, $\HUS^2\approx \HUS$ but $\HUS^2\HUS'$ is not associated to $\HUS \HUS'$: indeed $\HUS^2\approx \HUS$ implies by differentiation $2\HUS \HUS'\approx \HUS'$ and $\HUS^3\approx \HUS$ implies $3\HUS^2\HUS'\approx \HUS'$. Therefore, association is some kind of weak equality not coherent with multiplication and strictly weaker than the (new) equality $=$. 

Vector valued generalized functions can easily be introduced, but the (presumably very difficult) problems stated in the abstract precisely come from the fact that we have to deal with generalized functions whose values are (densely defined) unbounded operators on the Fock space (i.e., a specific Hilbert space used in QFT). In continuum mechanics and general relativity one could use scalar valued generalized functions, what was far simpler and fortunately permitted to solve the problems with just nonlinear generalized functions. In QFT we are confronted with considerably more complicated mathematics. However, by making clear the mathematical background, nonlinear generalized functions permit a mathematical statement of the remaining difficulties (unbounded operators, computer calculations, and renormalization theory).

\section{Difficulties in the physical interpretation 
         of nonlinear generalized functions}
\label{dif:0}

 Nonlinear generalized functions may have a (somewhat hidden) physical interpretation that is rather different from the one of classical functions. For instance they are not defined by pointvalues: two different generalized functions on an open set $\Omega$ in $\mathbb{R}$ may have the same pointvalues at all points of $\Omega$, \cite[p.\,37]{G-K-O-S}, see also \cite{K-S}. The standpoint (of distribution theory) that the interacting field operators have to be smeared over a finite space-time region is not abandoned when we shall define them at a fixed point $\bar{x} = (x,t)$ of space-time by means of a nonlinear generalized function $f$.  Indeed, because the ``pointvalue'' $f(\bar{x})$ involves an ``infinitesimal'' neighborhood of $\bar{x}$, the points have an ``infinitesimal'' but nonzero ``extension,'' which is made to tend to 0 in the definitions, presumably by ignorance of what it is in a physical context.

 Further, the physical interpretation of the word ``infinitesimal'' used above and the way to state physics with nonlinear generalized functions may depend on a scale (dimension, energy) of the phenomenon under consideration and may be based on an approximation at this scale: in \cite{C1,C2,C3,C8} the scale is the width of shock waves. Phenomena that have no classical analog occur in the $\mathcal{G}$-context even for solutions of simple partial differential equations (see \cite[Sec.\,4]{C2} for the appearance of a void creating a ``particle''). In short the concept of generalized functions compatible with nonlinear operations is extremely rich and subtle, without analogue in classical functions or distributions: there is an infinity of Heaviside step functions, of Dirac delta functions; generalized functions have an ``internal structure,'' i.e., for each $f$ each point $\bar{x}$ is in fact replaced by a portion of space-time (some kind of space-time ``quantization?'') that can be imagined as an approximation of a Dirac delta function centered at $\bar{x}$. Then $\bar{x}$ and $f$ cannot practically be dissociated, and the pointvalue $f(\bar{x})$ may even be related to the whole space-time,\footnote{Of course this ``dependence vanishes very rapidly at infinity,'' remaining essentially, but not completely, a point dependence.} thus introducing some very subtle kind of nonlocality, although at a rougher scale the locality expressed by the canonical commutation relations is fully satisfied... 

Therefore the underlying possible interpretations appear presently extremely numerous and one should avoid too hastly interpretations.  In QFT the phenomena under consideration are much more mysterious than in classical mechanics, so one should be very cautious with interpretations.\footnote{In contradistinction, the interpretation in \cite{C1,C2,C3,C4,C8} of the use of nonlinear generalized functions to model shock waves is rather clear because it reflects classical ideas of physics and it has been comforted by a number of calculations and experiments.}

\section{Aim of this paper} 
\label{aim:0}

This paper is focussed on the nonlinear calculations on distributions that constitute the basis of the Heisenberg-Pauli calculations to show that they can make sense mathematically. 

\emph{We only pretend to give a mathematical sense (very far from the`` best one'') to the H-P calculations as a starting point for improvement. We do not try, considering this would be premature, to interpret them physically. We stick mathematically to the H-P calculations in the simplest possible way: the interacting field operators will be usual functions of the time variable and generalized functions of the space variables because the time variable plays a privileged role in the (formal) H-P calculations.\footnote{While the H-P procedure with its asymmetrical treatment of the time is not manifestly relativistically invariant, it was shown by Heisenberg and Pauli that Lorentz-invariance is not destroyed by canonical quantization, what can also be verified by direct calculation, e.g., \cite{K}.}  Of course they should be presented in a next step as generalized functions on truly four-dimensional space-time.} 

\emph{Problem}: Modify the presentation below so that it becomes more suitable for the solution of the problems mentioned in the abstract, or for the physical interpretation. May be also consider more realistic physical models involving fermions.

\section{Prerequisites in QFT: The Fock space and the free-field operators}
\label{foc:0}

\subsection{Fock space}

The space of states called Fock space is the Hilbertian direct sum 
\begin{equation}\label{foc:1}
        \mathbb{F} = \bigoplus_{n=0}^\infty
                     \mathsf{L}_S^2\bigl((\mathbb{R}^3)^n \bigr), 
\end{equation}
where, for $n>0, \mathsf{L}_S^2\bigl((\mathbb{R}^3)^n \bigr)$ is the Hilbert space of complex valued symmetric square integrable functions on $(\mathbb{R}^3)^n$, with respect to the Lebesgue measure, and for $n=0, \mathsf{L}_S^2\bigl((\mathbb{R}^3)^n \bigr)$ stands for the field of complex numbers. That is: an element of $\mathbb{F}$ is an infinite sequence $(f_n)_n, n=0, ..., \infty$, such that $|f_0|^2 + \sum_{n=1}^{+\infty}(\|(f_n)\|_n)^2 < \infty$, where $\|~\|_n$ is the norm in $\mathsf{L}_S^2\bigl((\mathbb{R}^3)^n \bigr)$ and $f_n$ stands for the symmetric function $(\xi_1,... , \xi_n) \rightarrow f_n(\xi_1,... , \xi_n), \xi_i \in \mathbb{R}^3$.  Of course, from the definition of a Hilbertian direct sum, we have 
\begin{align}
\label{foc:2}
    \|(f_n)\|_\mathbb{F} &= \Bigl( |f_0|^2 + \sum_{n=1}^{+\infty}(\|(f_n)\|_n)^2 \Bigr)^{1/2},\\
\label{foc:3}
         \langle (f_n),(g_n) \rangle_\mathbb{F} &= f_0^* \cdot {g_0}
                     + \sum_{n=1}^{+\infty} \langle f_n,g_n \rangle_n,
\end{align}
where $\langle ~,~ \rangle_n$ is the scalar product in $\mathsf{L}_S^2\bigl((\mathbb{R}^3)^n \bigr)$ and $^*$ is complex conjugation.  In the sequel we shall use a dense subspace of $\mathbb{F}$, i.e.: 
\begin{equation}\label{foc:4}
     \mathbb{D} = \bigl\{~ (f_n)_n \in \mathbb{F}, 
                  \text{~~such that $f_n=0$ for $n$ large enough}~\bigr\}. 
\end{equation}

\subsection{Creation and annihilation operators}

If $\psi \in \mathsf{L}_S^2\bigl(\mathbb{R}^3\bigr)$ the creation operator $\mathbf{a}^+(\psi)$ is given by the formula: 
\begin{align}
\notag
      \mathbf{a}^+(\psi) ~:~  (f_n)_n \rightarrow
      \Bigl(& 0, f_0 \, \psi(\xi_1), ...,\\
\label{foc:5}
            &\sqrt{n} ~{\rm Sym}( \psi(\xi_n) \otimes f_{n-1}), ... \Bigr),
\end{align}
where ${\rm Sym}$ is the symmetrization operator. If $\psi \in \mathsf{L}_S^2\bigl(\mathbb{R}^3\bigr)$ the annihilation operator $\mathbf{a}^-(\psi)$ is given by the formula: 
\begin{align}
\notag
      \mathbf{a}^-(\psi) ~:~  (f_n)_n \rightarrow
      \Bigl(&\langle \psi(\xi_1), f_1(\xi_1) \rangle, ...,\\
\label{foc:6}
  &\sqrt{n+1}\langle \psi(\xi_{n+1}), f_{n+1}(\xi_1,..., \xi_{n+1}) \rangle,...
       \Bigr).
\end{align}
The operators $\mathbf{a}^+(\psi)$ and $\mathbf{a}^-(\psi)$ are defined at least on the dense subspace $\mathbb{D}$ of $\mathbb{F}$, with values in $\mathbb{D}$. They are not bounded operators on $\mathbb{F}$ because of the coefficients $\sqrt{n}$ and $\sqrt{n+1}$.

\subsection{Scalar product}

In the quantum theory of a `scalar' boson field the argument of the creation and annihilation operators is a Lorentz-invariant scalar function $\psi(\xi,t_\xi) \in \mathbb{C}$ where the variable $\{ \xi,t_\xi \} =  \{ \xi_\mu \} \in \mathbb{R}^4$ and $\mu = 1,..,4$.  The scalar product of two $1$-particle states is then not given by the usual $\mathsf{L}_S^2\bigl((\mathbb{R}^3)^n,\mathbb{C} \bigr)$ formula $\langle f_1,g_1\rangle = \int d^3\xi~f_1^*g_1$, and its obvious generalization to $n$-particle states, because the 3-volume element $d^3\xi$ is not Lorentz invariant.  Instead, it is the relativistically invariant expression \cite[p.\,828]{WIGHT1955-}, \cite[p.\,93]{K}
\begin{equation}\label{foc:7}
          \BRA  f_1(\xi) \| g_1(\xi) \KET
     :=  i \iiint_{t_\xi=\mathrm{Cst}} d^3\xi~\Bigl(
             f_1^* \frac{\partial g_1  }{\partial t_\xi} 
                 - \frac{\partial f_1^*}{\partial t_\xi} g_1 \Bigr),
\end{equation}
which is a special case of
\begin{equation}\label{foc:8}
          \BRA  f_1(\xi) \| g_1(\xi) \KET
     := i \iiint_\Sigma d^3\Sigma^\mu~\Bigl(
             f_1^* \frac{\partial g_1  }{\partial \xi_\mu} 
                 - \frac{\partial f_1^*}{\partial \xi_\mu} g_1 \Bigr),
\end{equation}
where $\Sigma$ is a space-like hypersurface, and $\mu$ a tensor index so that the contraction $\Sigma^\mu {\partial_{\xi_\mu}}$ is indeed invariant. Taking for $\Sigma$ the hyperplane  $t_\xi=\mathrm{Cst}$ leads to \eqref{foc:7}.

\subsection{Free-field operators}

The free-field operator is defined by: 
\begin{align}
\label{foc:9}
         {\bphi}_0(x,t)
           = \mathbf{a}^+\bigl(\Delta_+(\xi-x, t_\xi-t) \bigr) 
           + \mathbf{a}^-\bigl(\Delta_+(\xi-x, t_\xi-t) \bigr), 
\end{align}
where the variable $\{x,t\} \in \mathbb{R}^4$ corresponds to the space-time dependence of ${\bphi}_0$.  The function
\begin{equation}\label{foc:10}
  \Delta_+(x,t) = \frac{1}{(2\pi)^3} \int_{k \in\mathbb{R}^3}
                  \frac{d^3k}{2 k^0} \exp i (k \cdot x - k^0 t),
\end{equation}
where $k^0 = \sqrt{k^2+m^2}$ and $k \in \mathbb{R}^3$, is not in $\mathsf{L}_S^2\bigl(\mathbb{R}^3\bigr)$.  The mathematical object that makes sense is therefore:
\begin{equation}\label{foc:11}
              {\bphi}_0(\varphi,t) := \int_{\mathbb{R}^3}  {\bphi}_0(x,t) \varphi(x)~dx, 
\end{equation}
with $\varphi(x)$ a suitable test function.  That is: ${\bphi}_0(x,t)$ is a \emph{distribution} in the $x$-variable whose values ${\bphi}_0(\varphi,t)$ are densely defined linear unbounded operators on $\mathbb{F}$ (they map $\mathbb{D}$ into $\mathbb{D}$). The conjugate free-field operator is:
\begin{equation}\label{foc:12}
         {\bpi}_0(x,t) := \frac{\partial}{\partial t}{\bphi}_0(x,t), 
\end{equation}
which is again a similar distribution.  Of course it would be better, in view of Lorentz invariance, to consider them as distributions in $\{x,t\}$ but this is mathematically not necessary in order to stick to the H-P calculations.

\section{Prerequisites in QFT:  Sum-up of the Heisen\-berg-Pauli 
calculations in the simplest non-trivial case}
\label{sum:0}

In this section the Heisenberg-Pauli calculations (see for example \cite[p.\,21--22, 292--332]{W}) are summarized in the simplest non-trivial case, i.e., a self-interacting real scalar boson field. They are presented as purely formal calculations, i.e., they are not defined mathematically and are done by analogy with usual calculations on $\mathcal{C}^\infty$  functions. 

Let $\tau, g \in \mathbb{R}$. One sets \cite[p.\,206]{K}: 
\begin{align}
\notag
       \mathbf{H}^{(0)}(\tau) &= \int_{y \in\mathbb{R}^3} \Bigl\{
         \frac{1}{2} \bigl({\bpi}_0(y,\tau)\bigr)^2 
       + \frac{1}{2} \sum_{1 \leq \mu \leq 3}
         \bigl(\partial_\mu {\bphi}_0(y,\tau)\bigr)^2 \\
\label{sum:1}  
       &+ \frac{1}{2} m^2
         \bigl({\bphi}_0(y,\tau)\bigr)^2 
       + \frac{g}{N+1} \bigl({\bphi}_0(y,\tau)\bigr)^{N+1} \Bigr\} ~d^3y, 
\end{align}
where $\partial_\mu = \partial/\partial x_\mu$.  Of course this formula involves classically unjustified products of distributions whose values are unbounded operators on a Hilbert space, and also an unjustified integration. Then one sets: 
\begin{equation}\label{sum:2}
     {\bphi}(x,t,\tau)   =  \exp\bigl( i(t-\tau)\mathbf{H}^{(0)}(\tau)\bigr)
                              .{\bphi}_0(x,\tau).
                               \exp\bigl(-i(t-\tau)\mathbf{H}^{(0)}(\tau)\bigr).
\end{equation}
Again, the exponentials are not defined: they should be defined as operators (expected to be unitary) on $\mathbb{F}$, and there is also a problem of composition of operators because of the domain of ${\bphi}_0(x,\tau)$. Then calculations mimicking calculations on $\mathcal{C}^\infty$ functions give ${\bphi}(x,t,\tau)$ as the solution of the Cauchy problem called {\it ``interacting field equation:''} 
\begin{align}
\label{sum:3}
   \frac{\partial^2}{\partial t^2} {\bphi}(x,t, \tau) = 
   \sum_{1\leq\mu\leq 3} \frac{\partial^2}{\partial{x_\mu}^2} {\bphi}(x,t,\tau)
   - m^2 {\bphi}(x,t,\tau)
   - g \bigl({\bphi}(x,t,\tau)\bigr)^N,
\end{align}
with the initial conditions
\begin{align}
\label{sum:4}
      {\bphi}(x,\tau,\tau) = {\bphi}_0(x,\tau),
      \qquad \text{and} \qquad
      \frac{\partial}{\partial t} {\bphi}  (x,\tau,\tau) =
      \frac{\partial}{\partial t} {\bphi}_0(x,     \tau). 
\end{align}
It is a wave equation with nonlinear second member. Since the initial condition is a pair of irregular distributions the solution is not expected to be more regular than a distribution for which the nonlinear term does not make sense in distribution theory (with further a ``big'' problem due to the fact one is confronted with unbounded operators). Many such equations have been successfully studied in the $\mathcal{G}$-context, but in the scalar valued case, see the survey \cite{O2} and references therein. 

   To solve \eqref{sum:3} explicitly one introduces the operator:
\begin{align}\label{sum:5}
    \mathbf{S}_\tau(t) := \exp\bigl( i(t-\tau)\mathbf{H}_0\bigr)
                 \exp\bigl(-i(t-\tau)\mathbf{H}^{(0)}(\tau)\bigr), 
\end{align}
where $\mathbf{H}_0$ is the ``free-field Hamiltonian,'' i.e., Eq.~\eqref{sum:1} with $g=0$. The operator $\mathbf{S}_\tau(t)$ is considered as unitary on $\mathbb{F}$. Then, with,
\begin{equation}\label{sum:6}
     {\bphi}_I(x,t,\tau)   =  \exp\bigl( i(t-\tau)\mathbf{H}_0(\tau)\bigr)
                              .{\bphi}_0(x,\tau).
                               \exp\bigl(-i(t-\tau)\mathbf{H}_0(\tau)\bigr),
\end{equation}
one has (from calculations): 
\begin{align}\label{sum:7}
        {\bphi}(x,t,\tau)  = \bigl(\mathbf{S}_\tau(t)\bigr)^{-1}
                                .{\bphi}_I(x,t).
                                \mathbf{S}_\tau(t).
\end{align}
This formula suggests that the numerical results of the theory are the limits  when $\tau \rightarrow -\infty, t \rightarrow +\infty$ of the scalar products:
\begin{equation}\label{sum:8}
  \lim_{\substack{
       \tau \rightarrow -\infty\\
          t \rightarrow +\infty}} ~
 \bigl| \BRA \Phi_2 \| \mathbf{S}_\tau(t)\Phi_1\KET_\mathbb{F}\bigr|^2,
\end{equation}
where $\Phi_1, \Phi_2 \in \mathbb{D}$. One obtains (from calculations) that $\mathbf{S}_\tau(t)$ is solution of the differential equation: 
\begin{align}
\label{sum:9}
        \frac{\partial}{\partial t} \mathbf{S}_\tau(t) &=
        - i \frac{g}{N+1} \int_{x \in\mathbb{R}^3} 
        \bigl({\bphi}_I(x,t,\tau)\bigr)^{N+1} d^3x. \mathbf{S}_\tau(t),  \\
\label{sum:10}
        \mathbf{S}_\tau(\tau) &= \mathbf{1}, 
\end{align}
where $\mathbf{1}$ is the identity operator.  This formula is the starting point for an attempt (called \emph{``perturbation theory''}) to calculate $\mathbf{S}_\tau(t)$ by developing it in powers of the \emph{``coupling constant''} $g$, when $g$ is small.

{\it Remark:} These calculations form a rather intricate set of nonlinear calculations on distributions. Schwartz's impossibility result, see \cite{C1}, \cite[p.\,8]{C8}, and \cite[p.\,6]{G-K-O-S}, has been interpreted as the proof that these calculations cannot make sense even if we forget that they deal with unbounded operators on a Hilbert space (which makes them considerably more intricate than calculations in the case of bounded operators, close to scalar calculations). But it appears now that Schwartz's impossibility result is circumvented (Sec.\,\ref{gen:0}, \cite[...]{C1,C2,C3,G-K-O-S}), so that we may give a rigorous mathematical sense to these calculations, although with severe needs of improvements due the fact that they deal with unbounded operators.

\section{Preparation}
\label{pre:0}

If $\varphi$ is a suitable function on $\mathbb{R}^3$ we define its Fourier transform by: 
\begin{equation}\label{pre:1}
  (\mathsf{F}\varphi)(k) = (2\pi)^{-3/2} \int_{\mathbb{R}^3} \operatorname{e}^{-ik\cdot x} \varphi(x) ~dx.
\end{equation}
From now on $\varphi$ is such that its Fourier transform $\mathsf{F}\varphi$ is $\mathcal{C}^\infty$ with compact support in $\mathbb{R}^3$ and identical to 1 on a 0-neighborhood in $\mathbb{R}^3$. 

    In view of \eqref{foc:11} we first rewrite the free-field operator \eqref{foc:9} so that it reads:
\begin{align}
\label{pre:2}
     {\bphi}_0(\varphi,\epsilon,x,t)
       = \mathbf{a}^+\bigl(\Delta_\epsilon(\xi-x, t_\xi-t) \bigr) 
       + \mathbf{a}^-\bigl(\Delta_\epsilon(\xi-x, t_\xi-t) \bigr), 
\end{align}
where 
\begin{equation}\label{pre:3}
  \Delta_\epsilon(\varphi,\epsilon,x,t) =
          \frac{1}{(2\pi)^3}   \int_{k \in\mathbb{R}^3}
          \frac{d^3k}{2 k^0} \exp i (k \cdot x - k^0 t)
         .(\mathsf{F}\varphi)(+\epsilon k),
\end{equation}
and $\epsilon$ is a strictly positive real parameter that is arbitrarily close to 0.

   A further basic assumption will be that $\varphi$ is real valued.  Therefore ${\bphi}_0(\varphi,\epsilon,x,t)$ is a symmetric operator on $\mathbb{F}$. We set:   
\begin{align}\label{pre:4}
     {\bpi}_0(\varphi,\epsilon,x,t) = \frac{\partial}{\partial t}
                                         {\bphi}_0(\varphi,\epsilon,x,t).
\end{align}
${\bphi}_0(\varphi,\epsilon,x,t)$ and ${\bpi}_0(\varphi,\epsilon,x,t)$ are densely defined symmetric operators (of domain $\mathbb{D}$) on $\mathbb{F}$. They map $\mathbb{D}$ into $\mathbb{D}$ and satisfy on $\mathbb{D}$ the commutation relations (notation $[A,B]=AB-BA$): 
\begin{align}
\label{pre:5}
[{\bphi}_0(\varphi,\epsilon,x,t),{\bphi}_0(\varphi,\epsilon,x',t)] &=0,\\
\label{pre:6}
[ {\bpi}_0(\varphi,\epsilon,x,t), {\bpi}_0(\varphi,\epsilon,x',t)] &=0,\\
\label{pre:7}
[{\bphi}_0(\varphi,\epsilon,x,t), {\bpi}_0(\varphi,\epsilon,x',t)] &=
            i \delta(\varphi,\epsilon,x-x')~\mathbf{1}, 
\end{align}
where the nonlinear generalized function replacing the usual Dirac $\delta$-function is given by
\begin{equation}\label{pre:8}
        \delta(\varphi,\epsilon,x-x') = \frac{1}{\epsilon^3}
              (\mathsf{F}\rho)\Bigl(\frac{x-x'}{\epsilon}\Bigr),
\end{equation}
where $\rho$ is calculated from $\varphi$.  The function $\rho$ is $\mathcal{C}^\infty$ with compact support in $\mathbb{R}^3$, and is identical to 1 on a 0-neighborhood in $\mathbb{R}^3$. 

The above is equivalent to the classical description of the free fields considered as distributions, presented in a way convenient for the sequel. The formal description of physicists comes about by taking (abusively) $\epsilon=0$ (or by formulating the free-field operators by means of the nonlinear theory of generalized functions, which for the sake of clarity is postponed to Sec.\,\ref{int:0} where it will be really needed).

\section{Canonical Hamiltonian formalism: first step}
\label{fir:0}

We set: 
\begin{align}
\notag
  \mathbf{H}^{(0)}(\varphi,\FOU{\chi},\epsilon,\tau) &= \int_{y \in\mathbb{R}^3}
               \Bigl\{
      \frac{1}{2} \bigl({\bpi}_0(\varphi,\epsilon,y,\tau)\bigr)^2 
    + \frac{1}{2} \sum_{1 \leq \mu \leq 3}
      \bigl(\partial_\mu {\bphi}_0(\varphi,\epsilon,y,\tau)\bigr)^2 \\
\label{fir:1}  
    + \frac{1}{2} m^2
      \bigl({\bphi}_0&(\varphi,\epsilon,y,\tau)\bigr)^2 
    + \frac{g}{N+1} \bigl({\bphi}_0(\varphi,\epsilon,y,\tau)\bigr)^{N+1}
               \Bigr\}.\FOU{\chi}(\epsilon y) ~d^3y, 
\end{align}
with $\FOU{\chi}$ a $\mathcal{C}^\infty$ real valued function on $\mathbb{R}^3$ with compact support and identical to 1 in a 0-neighborhood.\footnote{$\FOU{\chi}(\epsilon y)$ is a damper introduced for the purpose of integrating over the whole of $\mathbb{R}^3$.}

    $\mathbf{H}^{(0)}(\varphi,\FOU{\chi},\epsilon,\tau)$ maps $\mathbb{D}$ into $\mathbb{D}$ and is symmetric. One can prove \cite[p.\,311--313]{C5} that it admits a self-adjoint extension denoted by $\mathbf{H}^{<0>}(\varphi,\FOU{\chi},\epsilon,\tau)$ on a domain denoted by $\mathbb{D}^{<0>}$ containing $\mathbb{D}$ (do not forget that $\mathbf{H}^{<0>}$ and $\mathbb{D}^{<0>}$ depend on $\varphi,\FOU{\chi},\epsilon,$ and $\tau$). Therefore, from the Hille-Yosida theory, $\{\exp(it \mathbf{H}^{<0>})\}_{t \in \mathbb{R}}$ is a strongly continuous group of unitary operators on $\mathbb{F}$. We set (considering $\tau$ fixed we do not note explicitly the $\tau$ dependence of ${\bphi}$, similarly we leave the dependence on $\FOU{\chi}$ implicit to simplify the notation): 
\begin{equation}\label{fir:2}
     {\bphi}(\varphi,\epsilon,x,t)   =
                 \exp\bigl( i(t-\tau)\mathbf{H}^{<0>}\bigr)
                  .{\bphi}_0(\varphi,\epsilon,x,\tau).
                 \exp\bigl(-i(t-\tau)\mathbf{H}^{<0>}\bigr),
\end{equation}
defined on
\begin{equation}\label{fir:3}
\mathbb{D}(t) := \bigl(\exp(i(t-\tau)\mathbf{H}^{<0>}\bigr)\mathbb{D},
\end{equation}
which is a dense subspace of $\mathbb{F}$ a priori depending on $t$ (and also on $\varphi,\FOU{\chi},\epsilon,$ and $\tau$).  We define ${\bpi}(\varphi,\epsilon,x,t)$ by formula \eqref{fir:2} with ${\bpi}_0$ in place of ${\bphi}_0$ (we do not yet know that ${\bpi}$ is the time derivative of ${\bphi}$). 

   Simplification of the exponentials and the commutation relations (\ref{pre:5} -- \ref{pre:7}) of the free-field operators give: 
\begin{align}
\label{fir:4}
 [{\bphi}(\varphi,\epsilon,x,t),{\bphi}(\varphi,\epsilon,x',t)] &=0,\\
\label{fir:5}
 [    {\bpi}(\varphi,\epsilon,x,t), {\bpi}(\varphi,\epsilon,x',t)] &=0,\\
\label{fir:6}
 [{\bphi}(\varphi,\epsilon,x,t), {\bpi}(\varphi,\epsilon,x',t)] &=
            i \delta(\varphi,\epsilon,x-x')~\mathbf{1}, 
\end{align}
which are valid on $\mathbb{D}(t)$.

\section{The problem of domains}
\label{dom:0}

From \eqref{fir:2}, ${\bphi}(\varphi,\epsilon,x,t)$ maps $\mathbb{D}(t)$ into $\mathbb{D}(t)$, and so does its expected time-derivative 
\begin{align}
\notag
               i \mathbf{H}^{<0>}.
                 \exp\bigl( i(t-\tau)\mathbf{H}^{<0>}\bigr)
                &.{\bphi}_0(\varphi,\epsilon,x,\tau).
                 \exp\bigl(-i(t-\tau)\mathbf{H}^{<0>}\bigr)\\
\notag
               + \exp\bigl( i(t-\tau)\mathbf{H}^{<0>}\bigr)
                &.{\bphi}_0(\varphi,\epsilon,x,\tau).
                 \exp\bigl(-i(t-\tau)\mathbf{H}^{<0>}\bigr)
           .(-i) \mathbf{H}^{<0>} \\
\label{dom:1}
               = i[ \mathbf{H}^{<0>}&,{\bphi}(\varphi,\epsilon,x,t)].
\end{align}
In order to define the $t$-derivative of ${\bphi}(\varphi,\epsilon,x,t)$ it would be desirable to have a domain independent of $t$, and this would be also desirable for the physical interpretation. (May be one should consider this ${\bphi}$ as generalized function on space-time for this problem of domain.)

{\it Problem:}  Find an extension ${\bphi}_{<0>}$ of ${\bphi}_0$ such that ${\bphi}$ defined by \eqref{fir:2} with ${\bphi}_{<0>}$ would have a dense domain $\mathbb{D}_1$ independent of $t$ (and also if possible of $\varphi,\FOU{\chi},\epsilon,$ and $\tau$), or find another way to have a domain independent of $t$. (May be consider generalized functions in the variable $(x,t)$ making Lorentz invariance evident.)

In the absence of such a $t$-independent domain we content ourselves with a weaker definition of the $t$-derivative. We say that $t \mapsto {\bphi}(t)$ from $\mathbb{R}$ to the set of linear operators from $\mathbb{D}(t) \rightarrow \mathbb{D}(t) \subset \mathbb{F}$ is differentiable with derivative $t \mapsto {\bfeta}(t)$ iff: for any $\mathcal{C}^1$ map $\Phi: t \mapsto \Phi(t)$ from $\mathbb{R}$ to $\mathbb{F}$, with $\Phi(t) \in \mathbb{D}(t)$ and $\Phi'(t)  \in \mathbb{D}(t), \forall t$, then the map $t \mapsto {\bphi}(t).\Phi(t)$ from $\mathbb{R}$ to $\mathbb{F}$ is differentiable with derivative $t \mapsto {\bfeta}(t).\Phi(t) + {\bphi}(t).\Phi'(t)$. With the above $\mathbb{D}(t)$ such a ${\bfeta}$ is unique.  Other definitions are possible to circumvent (at least provisionally) this domain problem. One obtains: 
\begin{align}
\label{dom:2}
     \frac{\partial}{\partial t} {\bphi}(\varphi,\epsilon,x,t)    \Phi
      = i[ \mathbf{H}^{<0>}, {\bphi}(\varphi,\epsilon,x,t) ]  \Phi,\\
\label{dom:3}
     \frac{\partial}{\partial t} {\bpi} (\varphi,\epsilon,x,t)    \Phi
      = i[ \mathbf{H}^{<0>}, {\bpi} (\varphi,\epsilon,x,t) ]  \Phi,
\end{align}
for any $\Phi$ in $\mathbb{D}(t)$.

\section{Canonical Hamiltonian formalism: second step}
\label{sec:0}

Set 
\begin{align}
\notag
    \mathcal{H}(\varphi,\epsilon,y,t) &=  \Bigl\{
      \frac{1}{2} \bigl({\bpi}(\varphi,\epsilon,y,t)\bigr)^2 
    + \frac{1}{2} \sum_{1 \leq \mu \leq 3}
      \bigl(\partial_\mu {\bphi}(\varphi,\epsilon,y,t)\bigr)^2 \\
\label{sec:1}  
    &+ \frac{1}{2} m^2
      \bigl({\bphi}(\varphi,\epsilon,y,t)\bigr)^2 
    + \frac{g}{N+1} \bigl({\bphi}(\varphi,\epsilon,y,t)\bigr)^{N+1}
               \Bigr\}. 
\end{align}
The compositions of these operators, from $\mathbb{D}(t)$ into $\mathbb{D}(t)$, make sense. Moreover, since $\mathbb{D}(t)$ is independent of $y$, the $\partial_\mu$  derivatives make sense. Simplification of the intermediate exponentials yields: 
\begin{align}\label{sec:2}
       \mathcal{H}(\varphi,\epsilon,y,t)  =
                 \exp\bigl( i(t-\tau)\mathbf{H}^{<0>}\bigr)
                  .\mathcal{H}^{(0)}(\varphi,\epsilon,y,\tau).
                 \exp\bigl(-i(t-\tau)\mathbf{H}^{<0>}\bigr),
\end{align}
where $\mathcal{H}^{(0)}$ is defined like $\mathcal{H}$ but with ${\bphi}_0$ and ${\bpi}_0$ instead of ${\bphi}$ and ${\bpi}$; $\mathcal{H}^{(0)}$ maps $\mathbb{D}$ into $\mathbb{D}$. Again simplification of the exponentials gives: 
\begin{align}\label{sec:3}
     \int_{y \in\mathbb{R}^3} 
               \mathcal{H}(\varphi,\epsilon,y,t).\chi(\epsilon y) ~d^3y
             = \mathbf{H}^{<0>}.
\end{align}
Thus (\ref{dom:2}--\ref{dom:3}) give: 
\begin{align}
\label{sec:4}
     \frac{\partial}{\partial t} {\bphi}(\varphi,\epsilon,x,t) ~\Phi
      = i \int_{y \in\mathbb{R}^3}
    [ \mathcal{H}(\varphi,\epsilon,y,t), {\bphi}(\varphi,\epsilon,x,t) ] 
         \chi(\epsilon y) ~d^3y~ \Phi,\\
\label{sec:5}
     \frac{\partial}{\partial t} {\bpi} (\varphi,\epsilon,x,t) ~\Phi
      = i\int_{y \in\mathbb{R}^3}
   [ \mathcal{H}(\varphi,\epsilon,y,t), {\bpi} (\varphi,\epsilon,x,t) ] 
         \chi(\epsilon y) ~d^3y~ \Phi,
\end{align}
if $\Phi \in \mathbb{D}(t)$. From (\ref{fir:4}--\ref{fir:6}) and the formula for $\mathcal{H}$ we get: 
\begin{align}\label{sec:6}
     \frac{\partial}{\partial t} {\bphi}(\varphi,\epsilon,x,t) ~\Phi
    &= i \int_{y \in\mathbb{R}^3}
         \delta(\varphi,\epsilon,y-x).{\bpi} (\varphi,\epsilon,y,t)
         \chi(\epsilon y) ~d^3y~ \Phi,\\
\notag
     \frac{\partial}{\partial t} {\bpi} (\varphi,\epsilon,x,t) ~\Phi
    &= i \int_{y \in\mathbb{R}^3}
      \Bigl\{ 
      \sum_{1 \leq \mu \leq 3} \partial_\mu {\bphi}(\varphi,\epsilon,y,t)
      .\partial_\mu \   \delta(\varphi,\epsilon,y-x)\\
\notag
    &+  m^2 {\bphi}(\varphi,\epsilon,y,t).\delta(\varphi,\epsilon,y-x)\\
\label{sec:7} 
    &+ g \bigl({\bphi}(\varphi,\epsilon,y,t)\bigr)^{N}
                 .\delta(\varphi,\epsilon,y-x)
      \Bigr\}.\chi(\epsilon y) ~d^3y~ \Phi,
\end{align}
if $\Phi \in \mathbb{D}(t)$. At this point we are forced to use the nonlinear generalized functions that we had avoided up to now to provide a presentation involving only classical objects.  These generalized functions provide a mathematical setting appropriate to take the limit $\epsilon \rightarrow 0^+$ so as to derive, in a suitable sense, the interacting field equation from (\ref{sec:6} --\ref{sec:7}).  They are also needed for the scattering operator.

\section{The interacting field operators}
\label{int:0}

This section is only a brief summary. 

On $\mathbb{D}$ we define a ``convenient kind of topological structure''  \cite[p.\,296]{C5}\footnote{We only say that a set in $\mathbb{D}$ is bounded iff it is contained in a finite sum in the Hilbert sum defining $\mathbb{F}$ and each of its components is bounded; these bounded sets play the role of a topology although there is no topology but this is a natural extension of the concept of a normed space.} and we define the algebra $\mathsf{L}(\mathbb{D})$ of linear operators from $\mathbb{D}$ into $\mathbb{D}$ that map any bounded set in $\mathbb{D}$ into (another) bounded set in $\mathbb{D}$; by definition a subset $M$ of $\mathsf{L}(\mathbb{D})$ is said to be bounded iff it is equibounded, i.e., $\forall B$ bounded set in $\mathbb{D}$ then $\{ L(x); L \in M, x \in B \}$ is again a bounded set in $\mathbb{D}$; $\mathsf{L}(\mathbb{D})$ becomes a ``bornological algebra''  in the terminology of \cite[p.\,286]{C5} . Then we define an algebra $\mathcal{G}\bigl(\mathbb{R}^3, \mathsf{L}(\mathbb{D})\bigr)$ of generalized functions whose values are linear operators from $\mathbb{D}$ into $\mathbb{D}$ \cite[p.\,287--295]{C5}. For fixed $t$, $\mathcal{G}\bigl(\mathbb{R}^3, \mathsf{L}(\mathbb{D}(t))\bigr)$ is defined by transport of structure through the exponentials that are in \eqref{fir:2}.  ${\bphi}(x,t)$ and ${\bpi}(x,t)$ are in $\mathcal{G}\bigl(\mathbb{R}^3, \mathsf{L}(\mathbb{D}(t))\bigr)$, as generalized functions of the variable $x \in \mathbb{R}^3$. 
\begin{quote}
{\it \underline{Theorem}.\footnote{The proof of this theorem is given in Ref.~\cite{C10}.} [Rigorous interacting-field equation]
\label{int:theo:1} 
In the context of operator-valued generalized functions in $\mathcal{G}$ one has, for all $\Xi(x) \in \mathcal{C}^\infty_0$, i.e., $\forall \Xi \in \mathcal{C}^\infty$ with compact support in the variable $x$,
\begin{align}
\label{int:1}
     \int d^3x~ \Xi(x) \Bigl\{
     \frac{\partial}{\partial t} \bphi  (x,t) 
 &-  \bpi(x,t) \Bigr\} = 0, \\
\nonumber 
     \int d^3x~ \Xi(x) \Bigl\{
     \frac{\partial}{\partial t} \bpi(x,t) 
 &-  \sum_{1\leq\mu\leq 3} \frac{\partial^2}{\partial{x_\mu}^2} \bphi(x,t)\\
\label{int:2}
 &- m^2  \bphi(x,t) - g \bigl(\bphi(x,t)\bigr)^N \Bigr\} = 0,
\end{align}
with the free-field operators as initial values at time $t=\tau$.
}
\end{quote}

{\it Translational and Lorentzian invariance}: Let $\mathcal{M}$ be the Minkowski space and let $\bar{x}, \bar{d} \in  \mathcal{M}$.  If $T(\bar{d})$ is a translation and $U(T(\bar{d}))$ its action on $\mathbb{F}$, then translation invariance consists of checking the formula ${\bphi}(\bar{x}+\bar{d})= U(T(\bar{d})).{\bphi}(\bar{x}).U(T(-\bar{d}))$. By following classical calculations, e.g., \cite{K}, it holds in the present context taking into account adequate changes in $\tau$ and $\chi$, even in the case of variable domains. At the end these changes disappear since $\tau \rightarrow -\infty$  and the function $y \rightarrow \chi(\epsilon y)$ covers the whole space as $\epsilon \rightarrow 0^+$. The same applies to Lorentz rotations preserving the direction of time \cite{K}. 

{\it Problem} (connected with the problem of domains): Improve the presentation so as to {\it put in evidence mathematically the properties of the interacting field operators} (hopefully they are ``formally'' correct since the above is a faithful transcription of the Heisenberg-Pauli calculations). In particular if $N=3$ formulate the properties listed in \cite[p.\,96--102]{WIGHT1964-} (Wightman axioms), of course in an adequate mathematical way (due to the generalized functions setting and the results obtained so far), but by {\it preserving their physical significance}.

\section{The scattering operator}
\label{sca:0}

We replace \eqref{sum:5} by:
\begin{align}\label{sca:1}
    \mathbf{S}_\tau(\varphi,\epsilon,t)
                       := \exp \bigl( i(t-\tau)\mathbf{H}_0^{<0>}\bigr)
                          \exp \bigl(-i(t-\tau)\mathbf{H}^{<0>}\bigr), 
\end{align}
where $\mathbf{H}^{<0>}(\varphi,\FOU{\chi},\epsilon,\tau)$ is the self-adjoint extension of the interacting-field Hamiltonian $\mathbf{H}^{(0)}$, and $\mathbf{H}_0^{<0>}$ that of the free-field Hamiltonian $\mathbf{H}_0$ which is equal to $\mathbf{H}^{<0>}$ when $g=0$. (To simplify the notation we leave implicit the dependence of $\mathbf{S}_\tau$ on $\FOU{\chi}$.)

$\mathbf{S}_\tau(\varphi,\epsilon,t)$ is a unitary operator on $\mathbb{F}$. Then one has (after calculations): 
\begin{align}\label{sca:2}
        {\bphi}(\varphi,\epsilon,x,t)  =
      \bigl(\mathbf{S}_\tau(\varphi,\epsilon,t)\bigr)^{-1}
    .{\bphi}_I(\varphi,\epsilon,x,t). \mathbf{S}_\tau(\varphi,\epsilon,t),
\end{align}
where ${\bphi}_I(\varphi,\epsilon,x,t)$ is the $\mathcal{G}$-version of \eqref{sum:6}, i.e.,  
\begin{equation}\label{sca:3}
 {\bphi}_I(\varphi,\epsilon,x,t) = \exp\bigl( i(t-\tau)\mathbf{H}_0^{<0>}\bigr)
                                   .{\bphi}_0(\varphi,x,\tau).
                                   \exp\bigl(-i(t-\tau)\mathbf{H}_0^{<0>}\bigr).
\end{equation}

 The crucial step is of course the calculation of the $t$-derivative of $\mathbf{S}_\tau(\varphi,\epsilon,t)$.  We do this according to the definition introduced in Sec.\,\ref{dom:0}.  This leads to the rigorous differential equation
\begin{align}\label{sca:4}
    \frac{\partial}{\partial t}
    \mathbf{S}_\tau(\varphi,\epsilon,t) 
        = -i \mathbf{H}_I(\varphi,\epsilon,t,\tau) ~ \mathbf{S}_\tau(\varphi,\epsilon,t),
\end{align}
where the $t$-dependent interaction Hamiltonian is now
\begin{align}
\label{sca:5}
    \mathbf{H}_I(\varphi,\epsilon,t,\tau)
   := \frac{g}{N+1} \int_{\mathbb{R}^3} d^3x~\FOU{\chi}(\epsilon x)
\bigl(\bphi_I(\varphi,\epsilon,x,t)\bigr)^{N+1}, 
\end{align}
in which the field operator $\bphi_I(\varphi,\epsilon,x,t)$ is defined by \eqref{sca:3}.  Note that \eqref{sca:5} is consistent because both $\mathbf{H}_I$ and $\bphi_I$ map $\mathbb{D}$ into $\mathbb{D}$.  Therefore, in \eqref{sca:4}, both $\mathbf{S}_\tau$ and ${\partial_t} \mathbf{S}_\tau$ map $\mathbb{D}(t)$ into $\mathbb{D}$.

   In combination with equations \eqref{sca:2} and \eqref{sca:3} the differential equation \eqref{sca:4} with initial condition $\mathbf{S}_\tau(\varphi,\epsilon,\tau) = \mathbf{1}$ solves the rigorous interacting field equation  (\ref{int:1}--\ref{int:2}) with the initial fields $\bphi_0(\varphi,\epsilon,x,t)$ and  $\bpi_0(\varphi,\epsilon,x,t) = \partial_t \bphi_0(\varphi,\epsilon,x,t)$ taken as those of a free field.  

   In the limits  $\tau \rightarrow -\infty, t \rightarrow +\infty$, which for now are formal limits, the operator $\mathbf{S}_\tau(\varphi,\epsilon,t) \equiv \mathbf{S}_\tau(\varphi,\FOU{\chi},\epsilon,t)$ becomes the \emph{scattering operator}, i.e.,
\begin{align}\label{scar:14}
    \mathbf{S}(\varphi,\FOU{\chi},\epsilon) := \lim_{t \rightarrow \infty}\mathbf{S}_{-t}(\varphi,\FOU{\chi},\epsilon,t),
\end{align}
which still depends on the mollifier $\varphi$, the damper $\FOU{\chi}$, and the parameter $\epsilon$.  The operator $\mathbf{S}_\tau(\varphi,\FOU{\chi},\epsilon,t)$ is unitary on $\mathbb{F}$, and gives the transition probabilities \eqref{sum:8}.  That is, as ``generalized real numbers,'' we have: 
\begin{align}\label{sca:6}
        \forall \Phi_1, \Phi_2 \in \mathbb{F}  \quad \Rightarrow \quad
        \bigl|\bigl\BRA\Phi_2 \| \mathbf{S}_\tau(\varphi,\FOU{\chi},\epsilon,t)
                          \Phi_1\bigr\KET_\mathbb{F}\bigr|^2, 
\end{align}
but calculating their limits as $\tau \rightarrow -\infty, t \rightarrow +\infty$, is non-trivial.

   Equation \eqref{sca:4} corresponds formally to \eqref{sum:9}, which leads to the (non-renorm\-alized) perturbation series in the standard formalism.

  {\it Problem:} Is it possible to calculate $\mathbf{S}_\tau$ perturbatively, i.e., as a formal series of powers of $g$, from \eqref{sca:4}?  Would something like the Epstein-Glaser method work \cite{E-G,Scharf}? 
Or else should one calculate globally starting from \eqref{sca:1}?

\section{The problems of calculating the transition probabilities}
\label{tra:0}

The transition probability that an initial state $\Phi_1 \in \mathbb{D}$ becomes after interaction a final state $\Phi_2 \in \mathbb{D}$ is given by something intuitively looking like the limits:
\begin{equation}\label{tra:1}
     \mathcal{P}(1 \rightarrow 2) =
       \lim_{\substack{
       \tau \rightarrow -\infty\\
          t \rightarrow +\infty}} ~
      |\BRA \Phi_2 \| \mathbf{S}_\tau(t)\Phi_1\KET_\mathbb{F}|^2.
\end{equation}
Before taking the limits these expressions are ``generalized real numbers'' between 0 and $\|\Phi_1 \|_{\mathbb{F}}^2 \cdot \|\Phi_2 \|_{\mathbb{F}}^2$  since the operator $\mathbf{S}_\tau(t)$ is a generalized unitary operator.  The problem is to extract an ``associated classical real number.''  The concept of association used till now in the context of nonlinear generalized functions consists of letting $\epsilon \rightarrow 0^+$ and finding a limit independent of the particular function $\varphi$ in use, i.e., here such that $\mathsf{F}\varphi$ is $\mathcal{C}^\infty$ with compact support in $\mathbb{R}^3$ and identical to 1 on a 0-neighborhood in $\mathbb{R}^3$.  Since it seems doubtful that such a limit would exist a more general definition of association is needed.

Therefore, a natural {\it probabilistic} interpretation of the association (of a classical number to a generalized number) is introduced by an averaging process: if $\bar{R}$ is a generalized number of representative $(\varphi,\epsilon) \rightarrow R(\varphi,\epsilon)$, intuitively bounded and oscillating as $\epsilon \rightarrow 0^+$ (think for example at $R(\varphi,\epsilon) = \exp (\cos (1/\epsilon))$ which endlessly oscillates between $1/{\rm e}$ and ${\rm e}$ as $\epsilon \rightarrow 0^+$) one may associate to $\bar{R}$ a mean value as $\epsilon \rightarrow 0^+$, provided it exists.  For instance, one may associate to the above $\bar{R}$ the value
\begin{equation}\label{tra:2}
  \lim_{\epsilon \rightarrow 0^+} \int_{0}^{1} \exp (\cos \frac{x}{\epsilon})~dx
  = \lim_{T \rightarrow \infty} \frac{1}{T} \int_{0}^{T} \exp (\cos x) ~dx,
\end{equation}
where the purpose of integrating over $[0,1]$ is just to capture the oscillations, as $\epsilon \rightarrow 0^+$, in order to take their average.  Other closely related averaging formulas, which give similar results, are possible.  We choose \eqref{tra:2} in the present context because it is the averaging formula considered in the theory of almost periodic functions \cite{Bes}.

   In $\mathbf{S}_\tau(\varphi,\epsilon,t)$ defined by \eqref{sca:1} one has from the theory of nonlinear generalized functions to let $\epsilon \rightarrow 0^+$, and further in \eqref{tra:1} one has to let $\tau \rightarrow -\infty$ and $t \rightarrow +\infty$, that is $t-\tau \rightarrow +\infty$.  It therefore appears that at least the $\epsilon$-limit should likewise be dealt with by an averaging process as above.  Since such an average always exists for all almost periodic functions one can ask:

  {\it Problem:} Is $|\langle \Phi_1, \mathbf{S}_{\tau,\epsilon}(t) \Phi_2\rangle|$ an almost periodic function in the variable $1/\epsilon$ (or possibly in the variables $t,\tau$ and $1/\epsilon$)?  And if not, does a mean value exist in a natural sense as  $\epsilon \rightarrow 0^+$ (or as $\epsilon \rightarrow 0^+$ and $t-\tau \rightarrow +\infty$)?

   From the theory of almost periodic functions, see \cite{Bes}, the answer is clearly yes for far simpler cases such as $|\langle \Phi_1, \exp(i\mathbf{A}/\epsilon) \Phi_2\rangle|$, where $\Phi_1,\Phi_2$ are in a finite dimensional Hilbert space and $\mathbf{A}$ is a symmetric operator. In our case we have to take into account the Hille-Yosida theory that is used to define $\exp\bigl(i(t-\tau)\mathbf{H}^{<0>}\bigr)$.

  {\it Problem:} Try to calculate numerically such an average.

\section{Final comments}
\label{fin:0}

We insist that the above should be considered only as an exercise on nonlinear generalized functions whose aim is to lay a groundwork for improvement --- that is, as a convenient model for presenting a novel mathematical framework to improve the mathematical consistency of quantum field theory.  

In particular, our discussion has been restricted to the major steps going from the Heisenberg-Pauli formulation of the scalar field problem to a simplified definition of the scattering matrix.  A comprehensive approach should discuss many pending problems related to the properties of the Green's functions, the definition of asymptotic states, the derivation of closed-form and perturbative solutions, renormalization, etc.  Moreover, the implications of nonlinear generalized functions should be related to the numerous efforts that have been made to give an axiomatic foundation to quantum field theory.

An application of the nonlinear theory of generalized functions as done above should be complemented by deeper works that rely on other mathematical theories as requested in problems. After 1955 QFT has been influenced by the claim that  ``there will never exist a theory of generalized functions compatible with nonlinear operations,''  which implied that the Heisenberg-Pauli calculations, although they have always been the backbone of QFT \cite{W}, would never make sense mathematically. Therefore the theory in \cite{C5,C6,C7} had to be first applied in {\it classical continuum mechanics and general relativity} (recent surveys \cite{C1,C2,C3,S-V}) in order to counterbalance this claim and be accepted by physicists as a {\it possible tool} in QFT, among other mathematical tools. 

The final conclusion is that the tool of nonlinear generalized functions --- which is rather ``soft'' but basic to have a well defined mathematical background as starting point --- appears to be convenient in that it opens doors by making mathematically understandable the overall framework of the Heisenberg-Pauli calculations. But, since QFT is very complicated, hard work using other needed mathematical tools such as the theory of unbounded operators on Hilbert spaces, and adequate computer calculations, are required before a final understanding can emerge. 

\section*{End note}
\label{end:0}

The detailed proofs of the calculations presented in this paper are given in Ref.~\cite{C10}.

\end{document}